# Π band dispersion along conjugated organic nanowires synthetized on a metal oxide semiconductor.

Guillaume Vasseur[*,1,2], Mikel Abadia[1], Luis A. Miccio[1,2], Jens Brede[*,1,2], Aran Garcia-Lekue[2,3], Dimas G. de Oteyza[1,2,3], Celia Rogero[1,2], Jorge Lobo-Checa[1,4,5] and J. Enrique Ortega[1,2,6]

[1.] *Centro de Física de Materiales (CSIC/UPV-EHU)-Materials Physics Center (MPC), Paseo Manuel Lardizabal 5, 20018 San Sebastián, Spain.*
[2.] *Donostia International Physics Center, Paseo Manuel Lardizabal 4, 20018 San Sebastián, Spain.*
[3.] *Ikerbasque, Basque Foundation for Science, 48011 Bilbao, Spain.*
[4.] *Instituto de Ciencia de Materiales de Aragón (ICMA), CSIC-Universidad de Zaragoza, 50009 Zaragoza, Spain.*
[5.] *Departamento de Física de la Materia Condensada, Universidad de Zaragoza, 50009 Zaragoza, Spain.*
[6.] *Departamento Física Aplicada I, Universidad del País Vasco, 20018 San Sebastián, Spain*

**ABSTRACT:** Surface-confined dehalogenation reactions are versatile bottom-up approaches for the synthesis of carbon-based nanostructures with predefined chemical properties. However, devices generally requiring low conductivity substrates, potential applications are so far severely hampered by the necessity of a metallic surface to catalyze the reactions. In this work we report the synthesis of ordered arrays of poly-p-phenylene chains on the surface of semiconducting $TiO_2$(110) *via* a dehalogenative homocoupling of 4,4"-dibromo-terphenyl precursors. The supramolecular phase is clearly distinguished from the polymeric one using Low Energy Electron Diffraction and Scanning Tunneling Microscopy as the substrate temperature used for deposition is varied. X-ray Photoelectron Spectroscopy of C 1s and Br 3d core levels traces the temperature of the onset of dehalogenation to around 475 K. Moreover, Angle-Resolved Photoemission Spectroscopy and Tight Binding calculations identify a highly dispersive band characteristic of a substantial overlap between precursor's π states along the polymer, considered as the fingerprint of a successful polymerization. Thus, these results establish the first spectroscopic evidence that atomically precise carbon-based nanostructures can readily be synthesized on top of a transition-metal oxides surface, opening the prospect for the bottom-up production of novel molecule-semiconductor devices.

## INTRODUCTION

The chemical versatility as well as relative abundance of carbon-based molecules makes them promising candidates for integration into next-generation devices such as molecular machines[1,2], organic field-effect transistors[3-5], or light emitting diodes[6,7]. However, technological applications require controllable, scalable, as well as cost efficient techniques before implementation into industrially manufactured devices becomes available. Toward this end, the exploitation of self-assembly of organic nanostructures on surfaces is one of the major challenges in the field of nanotechnologies[8-11]. In this respect, the surface-confined Ullmann reaction is a powerful bottom-up approach that exploits covalent bonding of halogenated precursor on top of metallic substrates[12-14]. The covalently bonded nanostructures exhibit superior thermal and mechanical stability, as well as improved charge transport properties that are mandatory in the field of electronics[15,16]. In addition, both geometry and electronic properties, including the band-gap, can be tailored by a judicious choice of monomers [17-20]. To date, a wide variety of these structures have been reported, including well-defined clusters[21], one dimensional wires[22-24], two-dimensional networks[25-27] and a large panel of width-controled graphene nanoribbons (GNR)[28-31]. However, practical applications are so far limited by the use of a metallic substrate, requisite to catalyze the reaction. Further processing is thus required to transfer the synthesized structures onto device-ready semiconductor surfaces which enable back-gating and minimize leak currents as well as substrate hybridization effects[32]. Recent experiments performed by local probe techniques indicate that the synthesis of covalently bonded nanostructures on bulk insulator surfaces[33-37], transition metal oxides[36,37], and also passivated semiconductor[38] may be feasible, but in particular the electronic properties of the synthesized structures have not been established yet.

Toward this end, photoemission spectroscopy is known as a powerful technique to investigate electronic states in self-assembled organic layers on surfaces. It is commonly used to probe the energies of molecular levels, thereby allowing tracing such properties as molecule-substrate interactions[39] or changes in the molecular chemical composition[40]. Moreover, the angular dependence of photoemission intensity contains rich information about orbital topologies[41]. Using inverse Fourier transformation these orbitals can be recovered[42], including their phase which is supposed to be lost during the photoemission process[43]. Band dispersions were also reported for molecular layers and crystals as the signature of weak electronic coupling mediated either directly by intermolecular interations[44] or indirectly by the substrate[45]. Direct evidence of conjugation associated with covalent bonding was reported for bottom-up synthesized carbonaceous materials such as 2D covalent networks[46] and graphene nanorribons[47]. More recently, the complete valence band structure of poly-p-phenylene (PPP) has been measured both on Cu(110)[48] and Au(788)[49]. We note that the long range ordering as well as uniaxial alignment requisite for angle resolved photoemission measurements of one dimensional systems is frequently achieved exploiting inherently anisotropic substrates such as high index surfaces[47,49,50].

Herein we report the first spectroscopic study of ordered arrays of PPP chains directly synthesized on top of a titanium dioxide surface. The growth of 4,4"-dibromo-terphenyl (DBTP) precursors is followed by Low Energy Electron Diffraction (LEED), Scanning Tunneling Microscopy (STM), and X-ray photoemission spectroscopy (XPS) as a function of substrate temperature. LEED and STM clearly resolve two distinct and well-ordered structural phases with a transition temperature between the phases of about 475 K. This structural transition coincides with a shift of both C 1s and Br 3d core levels, as observed in XPS. Similar shifts are often considered as the fingerprint of the dehalogenation of precursors. Successful polymerization is established by angle resolved photoemission (ARPES), resolving the transition from a single, well defined Highest Occupied Molecular Orbital (HOMO), corresponding to a supra-molecular phase, to a continuous valence band dispersion associated with the PPP polymer.

## RESULTS AND DISCUSSION

**Structural characterization.** The rutile-TiO$_2$(110) 1x1 substrate (Figure 1.a-c) consists mainly in alternating rows of fivefold-coordinated titanium (5f-Ti) and twofold- or threefold-coordinated oxygen atoms (2f-/3f-O), running along the [001] direction[51]. Due to its rectangular unit cell (2.95 x 6.49 Å), the surface is characterized in LEED by an anisotropic pattern, presented in figure 1.a. This inherent anisotropy is further enhanced by the 2f-O atoms protruding out of the surface plane, framing quasi one-dimensional trenches of fivefold coordinated 5f-Ti atoms as depicted in the model in Figure 1.c. The STM image (figure 1.b) of the surface is, at the given tunneling conditions, dominated by electronic rather than geometrical effects, hence, bright rows are assigned to 5f-Ti atoms and dark spaces in between to 2f-O ones[52]. Moreover, after reduction of the crystal by several sputtering and annealing cycles, the surface exhibits defects, in particular oxygen vacancies (Ovs, green arrows on figure 1.b,c), which are known to be involved in several catalytic properties of titanium dioxide such as water dissociation[53].

Upon evaporation of DBTP on top of the surface held at room temperature (RT) several additional stripes, highlighted by red arrows in Fig. 1d, appear in the LEED image. The separation between these stripes indicates a super-periodicity along the [001] direction of 2.1±0.2 nm. With saturation of the surface (see supplementary figure S1), the STM images exhibit well defined rows running along the [001] direction (figure 1.e). The distance between these rows is 6.5 ±0.1 Å, equal to the lattice parameter of the clean substrate in the [$\bar{1}$10] direction. Higher resolution data (figure 1.e bottom and figure S2) has revealed an intra-chain super-periodicity of 1.8 ±0.2 nm along the [001] axis, close to the value deduced in LEED. Furthermore, we no longer observe the characteristic O vacancies. All these considerations strongly suggest a one-dimensional arrangement of the DBTP molecules into supramolecular chains, as schematically depicted in Fig. 1e. This model can readily be rationalized considering density functional theory (DFT) calculations of the free standing molecules (Supplementary Figure S3, Figure S4 and Note 1). Geometry optimization leads to a molecule's length (Br to Br) of about 15.2 Å. Considering two molecules facing each other along their main axis, the Van-der-Waals interactions stabilize the inter-molecular Br-Br distance at around 3.8 Å. Therefore, the periodicity of a one dimensional DBTP supramolecular chain is about 1.9 nm, in agreement with the values experimentally deduced from LEED and STM. Note that in this model, the lack of a phase relation between two adjacent chains explains the apparition of a striped pattern instead of well-defined diffraction spots in LEED (Supplementary Figure S5 and Note 2).

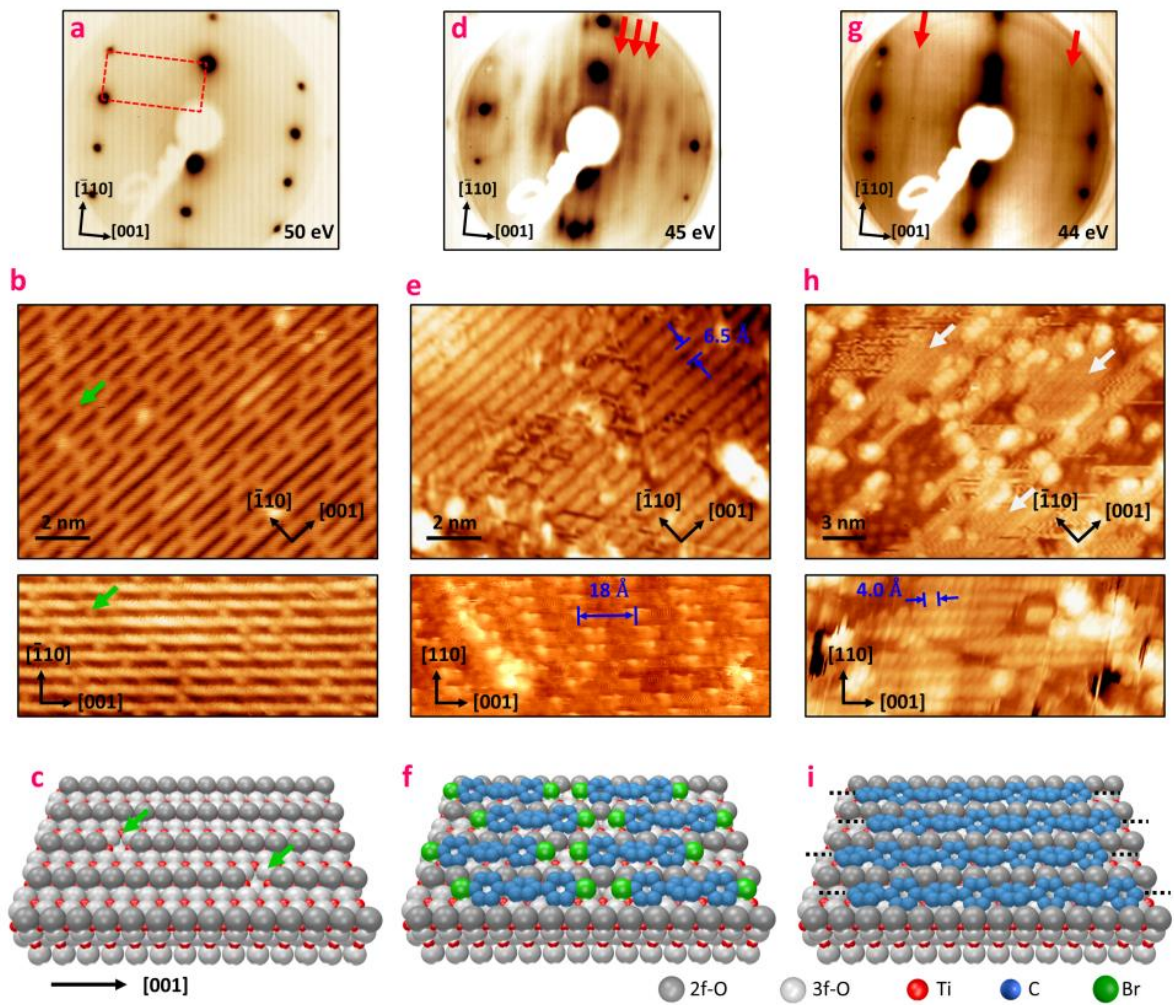

**Figure 1: Structural characterization of supramolecular and polymeric phases.** LEED patterns (a,d,g), room temperature STM images (b,e,h) and Ball models (c,f,i) associated to the clean rutile-TiO2(110) substrate (a,b,c), the supramolecular layer (d,e,f) and the polymeric layer (g.h.i). (d,e) have been measured after saturation of the surface with DBTP deposited at room temperature, (g) after saturation with the substrate held at 575 K, and (h) after semi-saturation with the surface held at 575 K. For clarity, hydrogen atoms have been omitted in the ball models. On (b) and (c), green arrows indicate some oxygen vacancies. The white arrows in (h) indicate the polymeric chains.

Post-annealing the sample prepared at room temperature leads to a clear transition characterized in LEED by the rapid disappearance of the supramolecular features around 470 K (Supplementary Figure S7). Only the characteristic spots of the substrate remain and no signature of any ordered molecular layer is observed in LEED. However, STM images (Supplementary Figure S8) show some short chain-like features oriented along the [001] direction and a multitude of disordered protrusions. While it is appealing to attribute the chain-like features to polymerized molecules[33-37], this assignment remains ambiguous without further characterization. To improve the overall order of the surface we proceeded to evaporating the precursors on the surface kept at 575 K. This procedure leads to the appearance of two well-defined stripes in LEED images, indicated by red arrows in figure 1.g. The corresponding lattice parameter along the [001] direction is about 4.2±0.2 Å. This value matches the theoretical inter-phenyl distance of 4.23 Å obtained by DFT for a PPP chain[54], as well as the one of 4.4 Å measured experimentally on Cu(110)[55]. STM images of a non-saturated layer (Supplementary Figure S1) reveal several chains running along the [001] direction, marked by white arrows on Fig. 1.h (see also Supplementary Figure S8.b). In this case, the bare $TiO_2$ surface is also readily identified by the familiar stripe pattern and the characteristic defects. In the [$\bar{1}$10] direction, surface and molecular chains show the same perpendicular spacing of 6.5 ±0.2 Å as the bare substrate and the supramolecular layer. An intra-chain periodicity of 4.0 ±0.3 Å is deduced from high resolution STM images, which is in complete agreement with the formation of PPP chains running along oxygen rows, as depicted in figure 1.i. Once again, the lack of phase relation between two adjacent polymers is responsible for the striped pattern observed in LEED (Supplementary Figure S6). However, we note that up to now, for both supramolecular and polymeric phases, it has not been possible to determine with accuracy the absolute positions of the chains with respect to the substrate.

**Chemical characterization.** XPS analysis is intensively used to follow the evolution of core levels in several chemical reactions that occur on top of surfaces[56-58]. Especially on metals, the Ullmann coupling was recently studied both theoretically[14] and experimentally[59,60], revealing an organometallic intermediate that appears close to room temperature on copper and silver substrates. In our case, both C1s and Br3d core levels were followed as a function of sample temperature, in order to characterize the chemical process of the reaction on titanium dioxide.

Firstly, DBTP was evaporated on the sample kept at low temperature (LT, 80K) in order to characterize a multilayer of intact (non-debrominated) molecules (figure 2.a, bottom). In the Br3d region, two contributions associated to $Br3d_{3/2}$ and $Br3d_{5/2}$ are resolved with binding energies (BE) of 70.6 eV and 71.6 eV, respectively. Likewise, the spectrum in the C1s region is fitted by two contributions at 284.7 eV and 285.3 eV, respectively associated to the active and hydrogen passivated carbon atoms, in agreement with the literature[14,59,60]. After 1ML deposition at room temperature, both of C1s and Br3d features remain globally unchanged compared to LT multilayers and no significant shifts were observed (figure 2.a, middle). This result suggests that the molecules are still intact on the surface, in agreement with the model proposed in figure 1.f. However, for high temperature depositions, a substantial shift of both Br3d and C1s core levels, respectively by 1.4 eV and 0.2 eV towards lower BE, are observed. The shift of the Br3d doublet is attributed to the scission of the C-Br bond and the formation of Br-Ti species. Its magnitude of 1.4 eV is comparable to the theoretically predicted value of 1.6-1.9 eV core level shift on noble metal surfaces[14]. The evolution of the carbon line shape is in agreement with the formation of a polymeric phase, i.e. a C-C homocouplings of the dehalogenated precursors[59].

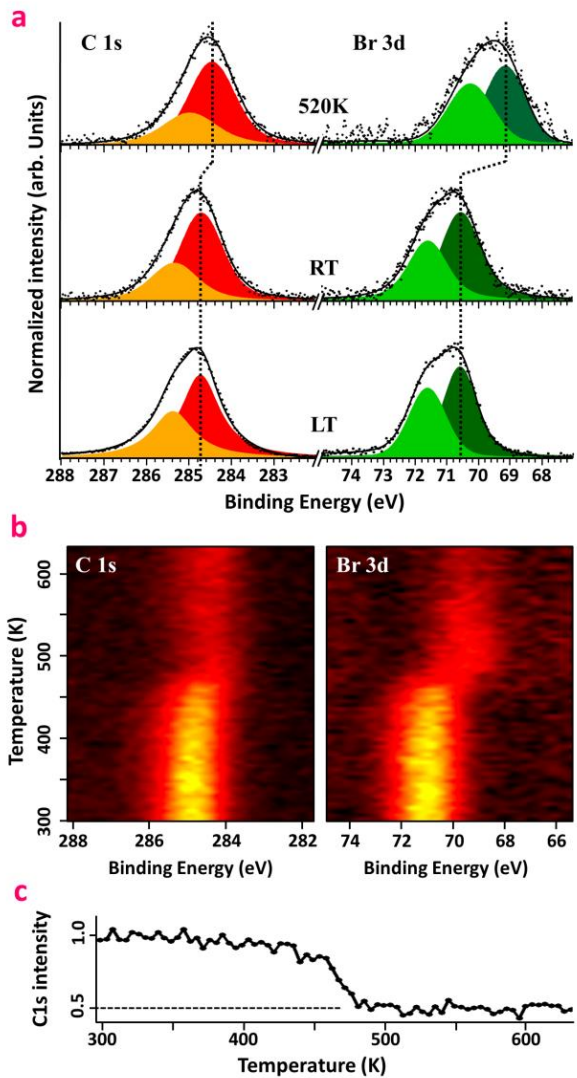

**Figure 2: X-ray Photoemission Spectroscopy;** (a) C1s and Br3d core level spectra measured after a multilayer deposition of DBTP with the sample kept at 80K (bottom), after 1ML deposition at RT (middle) and 1 ML at 520 K (top); (b) Evolution of C1s and Br3d core levels as function of temperature, measured after deposition of 1 ML at RT; (c) Evolution of the integrated C1s signal as function of temperature, extracted from (b).

A precise determination of the activation temperature for the high temperature (HT) phase formation is obtained by following the XPS spectra during the annealing process of a sample prepared initially at RT (Figure 2.b). The line shape of the C1s and Br3d core levels remains unchanged until 450K. Then, a clear shift accompanied with a substantial loss in photoemission intensity takes place within 50 K until about 500K. The obtained energy shifts are identical to the ones discussed above, but the reduction of intensity is about 50% throughout the transition as is depicted in Figure 2.c. This overall decline is distinct from previous experiments reported on metals suggesting that, due to the weakly interacting nature of the molecular precursor on the substrate, a large part of the molecules are desorbed during the annealing process. Since the intensity decrease coincides with the onset of the core levels shifts, we can conclude that the activation energy of C-Br-scission on $TiO_2$ is of the same order as the binding energy of DBPT molecules in the RT phase. Then, by increasing the temperature past 575 K, bromine is gradually desorbed from the surface. We notice that XPS data of a post-annealed RT-phase and a HT deposited phase are indistinguishable within our experimental resolution, i.e., dehalogenation as well as C-C coupling occur in both cases, even if the global order is improved after a high temperature deposition (Supplementary Figure S8). Furthermore, we exclude the possibility that the clusters observed with STM after polymerization are due to Br atoms remaining on the surface since these features are visible also in STM data taken on samples that showed no trace of the Br3d core level in XPS after annealing at higher temperature.

In summary, the XPS characterization shows all characteristic features of a dehalogenation reaction. However, in contrast to noble metal catalyzed reactions reported previously, a substantial loss of molecular precursors due to desorption from the surface is observed.

**Electronic properties.** The structural and chemical characterization of the RT and HT phases offers compelling evidence for the polymerization reaction. However, measurement of electronic dispersion throughout the formed PPP oligomers is still lacking. In many ways the band structure of the polymer is its key property and may be considered as its defining fingerprint. In particular, a large precursor's orbitals overlap associated to covalent bonding induces generally highly dispersive bands[47], while a small dispersion is rather a sign of weak interactions[44,48]. In order to probe the electronic levels in the different structural and chemical phases we directly mapped the electron dispersion with ARPES along the [001] direction, parallel to the oxygen rows. Incidents photons are p-polarized, with an energy of 21.22 eV. The comparison between measurements of the pristine substrate with the RT and HT molecular layers enables us to determine the features originating from molecules.

The clean substrate (figure 3.a and 3.d, grey trace) is characterized by broad features appearing between 3.3 and 9 eV below the Fermi energy, which correspond to the valence bulk bands (bVB) with a predominantly $O_{2p}$ character[61]. The ARPES map shows also a characteristic defect state (DS) close to 1eV, indicated in figure 3.d by the red arrow. The origin of this feature is still under debate and has been assigned both to O vacancies[62] and interstitial Ti atoms[63]. After deposition of 1 ML DBTP at room temperature, a well-defined molecular state appears just over the bVB, around -2.9 eV, as depicted in figure 3.b and 3.d (blue trace). Such levels have already been evidenced for a large variety of organic molecules on metallic surfaces[39,42] and are usually associated to their π-conjugated HOMO state. The photoemission intensity of this feature is modulated along the [001] direction, with two maxima appearing at 0.6 and 1.4 Å$^{-1}$. This spectral weight distribution, already observed for similar molecules such as sexiphenyl[42,44], is intrinsically related to the topology of the HOMO state and its Fourier transform[41-44]. Thus, the main contribution at 1.4 Å$^{-1}$ is directly related to the inter-phenyl distance (4.4 Å), while the one at 0.6 Å$^{-1}$ is due to the natural twisted conformation of the molecule[64] (see also supplementary Figure S3). Furthermore, a strong suppression of the defect state is also observed and will be discussed in the following section. Then, depositing DBTP at high temperature leads to the complete disappearance of the single molecular level, replacing them by a strongly dispersive band, as presented in figure 3.c and 3.d (green trace). The top of the band is located at -2.06 eV, shifted towards the Fermi energy by 0.9 eV when compared to the supramolecular phase. This shift is fully compatible with the HOMO/LUMO gap reduction expected for the formation of covalent C-C bonds between the dehalogenated precursors[46]. The band apex is reached at 1.45 Å$^{-1}$, in agreement with the polymer periodicity of 4.2±0.2 Å deduced by LEED. From the parabolic fit, we determine an effective mass of -0.17 $m_0$, with $m_0$ being the free electron mass. Another important quantity is the group velocity, which reaches a value of 9.2 x 10$^5$ m·s$^{-1}$ in the linear part of the band, around -1 eV below the maximum. Such characteristics are in perfect agreement with previous works reported on PPP oligomers[48,49], reflecting the large overlap between the precursor's molecular orbitals, and are more generally expected for graphene based materials like nanoribbons[47].

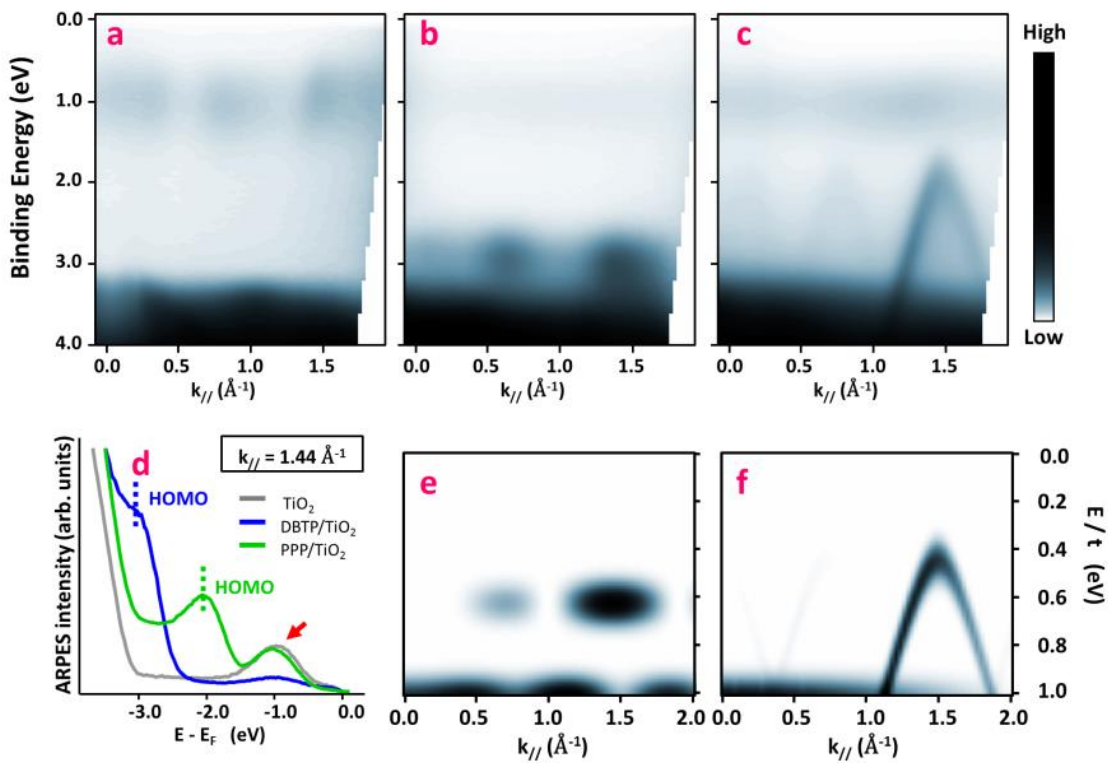

**Figure 3: Angle Resolved Photo Emission Spectroscopy;** (a-c) Experimental raw ARPES intensity maps acquired on the pristine surface (a), after 1ML DBTP deposition at room temperature (b), and after 1ML deposition with the sample kept at 575K (c); (d) Energy dispersion curves taken at $k_{//}$=1.44 Å$^{-1}$ from the three previous maps; (e,f) Theoretical ARPES intensity distributions calculated within a Tight-Binding model for a terphenyl molecule (e) and an infinite poly-p-phenylene chain (f). The energy scale (E) is normalized by the resonance integral (t) between the $p_z$ orbitals of two neighboring carbon atoms.

A simple tight-binding (TB) model allows us to qualitatively reproduce the main characteristics of the supramolecular and polymeric features observed in ARPES. The theoretical signature of the HOMO of a terphenyl molecule is depicted in figure 3.e. The state is characterized by two maxima appearing at 0.6 and 1.4 Å$^{-1}$, in agreement with experimental observations. Considering an infinite polymer chain, we observe that the discrete molecular states transform into a dispersive band, as depicted in figure 3.f. For an inter-phenyl distance of 4.4 Å, the top of this band is found at 1.43 Å$^{-1}$. Quantitatively, considering a resonance integral (t) between the two $p_z$ orbitals of neighboring C atoms equal to 2.7 eV (expected for graphene-like materials[65]), these calculations lead to an effective mass of -0.14 $m_0$, close to the one deduced experimentally. Therefore, we consider the observed band as the spectroscopic fingerprint of a successful polymerization of DBPT molecules into an ordered array of long PPP chains.

**Evolution of the defect peak in ARPES.** ARPES also provides important information about the mechanism responsible for the dehalogenation reaction occurring on titanium dioxide. As stated previously, for the clean surface, the reduction of the substrate by sputtering/annealing cycles generates defects, leading to an excess of charge that originates the Ti3d DS, which is visible in UPS around 1.0 eV below the Fermi energy (red arrow in figure 3.d). After RT deposition, this peak almost vanishes, but is fully recovered for the polymeric phase. In a recent publication, Kolmer *et al.* demonstrate on the (011) surface that the reaction is catalyzed by hydroxyl groups[37]. However, our STM images of the clean surface suggest only a negligible amount of hydroxyl groups prior to molecule deposition[66]. Although a small amount of hydroxyl groups may be created during thermal evaporation of DBTP onto the surface, the temperature of more than 575 K during molecule deposition for the polymerized HT-phase is sufficient to induce diffusion of H from the surface hydroxyl groups into the bulk of $TiO_2$(110)[67]. Detailed information on the rates of surface hydroxyl group annihilation and creation are thus necessary to conclude about their role in the present polymerization reaction, but other factors appear to play a role as well. In particular, the observed attenuation of the DS peak in our ARPES data was shown previously to not be related with the formation of OH groups[63]. Moreover, several other studies reported that the DS is preserved after thin layer depositions of different organic molecules[68] like Phthalocyanines[69] or Perylene-tetracarboxylic-dianhydride[70]. Thus, a direct screening of the DS by an organic layer seems very unlikely. Therefore, another interpretation of the ARPES data is the direct interaction of Halogen atoms with Ovs leading to the disappearance of the DS. A similar effect was observed previously for chlorine deposited[71] on $TiO_2$(110). However, in the present case question arises whether the quenching of the DS is due to a minute amount of Br atoms that were directly evaporated from the molecular source or whether Br-terminated molecules are interacting directly with the DS. Unfortunately, the resolution of our laboratory XPS source does not allow deciding between these two scenarios. Nonetheless, we can conclude that upon heating the sample during polymerization of DBTP a re-evaporation of Br correlates with the restoration of the DS.

## CONCLUSION

Here, we have presented a spectroscopic study of an *in-situ* polymerization of DBTP precursors into ordered arrays of poly-p-phenylene nanowires on top of the $TiO_2$(110) surface. The signatures of the DBTP supramolecular assembly as well as PPP oligomers were characterized by LEED, STM, XPS, ARPES, and TB calculations. In particular, LEED and STM data resolved long organic wires perfectly aligned along the Oxygen rows in the [001] crystal direction. Collecting temperature dependent XPS spectra, we have traced the C-Br bond scission temperature of the DBPT precursors which is accompanied with the onset of the polymerization process to 475 ± 25 K. Angle-resolved photoemission spectroscopy was used to map a strongly dispersive π-conjugated band upon formation of long PPP chains. Analysis of the data has revealed an effective mass of -0.17 $m_0$ at the top of the valence band as well as a group velocity of up to 9.2 x $10^5$ m/s in the linear part of the dispersion. The latter reflects the large orbital overlap originating from covalent coupling of the precursors and generally expected for conjugated systems. Therefore, the presented on-surface bottom-up synthesis opens up the prospect to the fabrication of atomically controlled nanostructure, such as graphene nanoribbons, on top of transition metal oxides. This paves a new pathway towards the integration of these structures into multifunctional electronics devices that will take advantage of both molecular and substrate properties.

## METHODS

Experiments were carried out in ultrahigh vacuum systems at base pressures of $10^{-10}$ mbar. Commercial $TiO_2$(110) single crystals (Cyrstek) were prepared by repeated cycles of sputtering ($Ar^+$,1 keV) and annealing (900 K). While still considered as a semiconductor[72], optical inspection of the crystals revealed a dark blue colour indicating a certain degree of reduction of the $TiO_2$ bulk. 4,4"-dibromo-terphenyl molecules (Sigma Aldrich) were initially degassed thoroughly for several hours/days by heating the source to temperatures slightly below 350K under UHV conditions and subsequently sublimed using a Knudsen cell heated up to about 375K, in order to obtain a rate close to 0.33 ML/min. For both RT and HT deposition, the sample saturates at 1ML, leading to a self-limited growth (Supplementary Figure S1)

**Scanning tunnelling microscopy** was carried out at RT using either an Omicron VT-STM or a SPECS Aarhus STM. STM images were recorded at constant tunnelling current (30 pA) and constant bias voltage (applied to the sample) between 1.8 and 2.5V.

**XPS** experiments were performed using a Phoibos photoelectron spectrometer equipped with an Al Kα X-ray source (16 mA, 12.5 kV) as the incident photon radiation. The overall resolution of the instrument is approximately 0.9 eV. During the temperature dependent measurement the temperature was increased at a rate of 1.7 K/min.

**Photoemission** measurements were performed using a Phoibos 150 SPECS high-resolution hemispherical electron analyser. The sample was cooled down to 150K. He-I (hν = 21.2 eV) radiation was provided by a high intensity UVS-300 SPECS discharge lamp coupled to a TMM-302 SPECS monochromator.

**Theoretical ARPES intensity** distributions where obtained by taking the square modulus of the Fourier transform of the molecular orbitals calculated for both Terphenyl and long PPP chains within the Tight-Binding model, according to the method presented in refs. 41 and 42.

## ACKNOWLEDGMENTS

This work was supported by the Spanish Ministry of Economy (grants MAT2013-46593-C6-4-P), the Basque Government (grants IT-621-13), and the European Research Council under the EU Horizon 2020 research and innovation programme (grant agreement No 635919). We acknowledge also support from the Basque Department of Education, UPV/EHU (Grant No. IT-756-13), the Spanish Ministry of Science and Innovation (MAT2013-46593-C6-2-P), and the European Union FP7-ICT Integrated Project PAMS (Contract No. 610446).

## ASSOCIATED CONTENT

**Supporting Information**. Self-limited growth of supramolecular and polymeric phases, additional STM images of the supramolecular phase, geometry optimization of the DBTP molecule and supramolecular phase, theoretical supramolecular and polymeric LEED patterns, phase transition in LEED, comparison of post-annealed room temperature preparation with high temperature deposition. "This material is available free of charge via the Internet at http://pubs.acs.org."

## AUTHOR INFORMATION

Corresponding Authors
*E-mail: g.vasseur.univ@gmail.com
*E-mail: dr.jens.brede@gmail.com

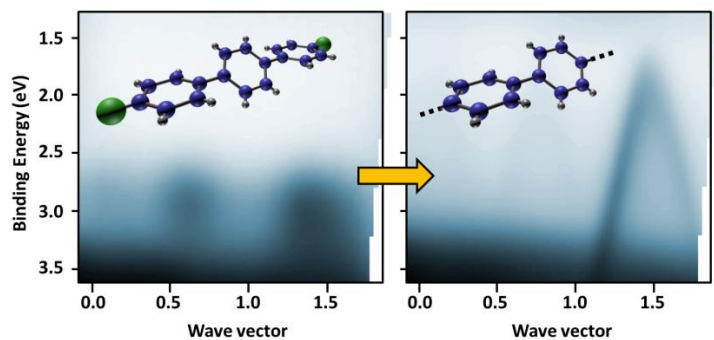

**Table of contents (TOC)**

# Supplementary Information

# Π band dispersion along conjugated organic nanowires synthetized on a metal oxide semiconductor.

Guillaume Vasseur[*,1,2], Mikel Abadia[1], Luis A. Miccio[1,2], Jens Brede[*,1,2], Aran Garcia-Lekue[2,3], Dimas G. de Oteyza[1,2,3], Celia Rogero[1,2], Jorge Lobo-Checa[1,4,5] and J. Enrique Ortega[1,2,6]

[1.] *Centro de Física de Materiales (CSIC/UPV-EHU)-Materials Physics Center (MPC), Paseo Manuel Lardizabal 5, 20018 San Sebastián, Spain.*
[2.] *Donostia International Physics Center, Paseo Manuel Lardizabal 4, 20018 San Sebastián, Spain.*
[3.] *Ikerbasque, Basque Foundation for Science, 48011 Bilbao, Spain.*
[4.] *Instituto de Ciencia de Materiales de Aragón (ICMA), CSIC-Universidad de Zaragoza, 50009 Zaragoza, Spain.*
[5.] *Departamento de Física de la Materia Condensada, Universidad de Zaragoza, 50009 Zaragoza, Spain.*
[6.] *Departamento Física Aplicada I, Universidad del País Vasco, 20018 San Sebastián, Spain*

* Corresponding authors
g.vasseur.univ@gmail.com
dr.jens.brede@gmail.com

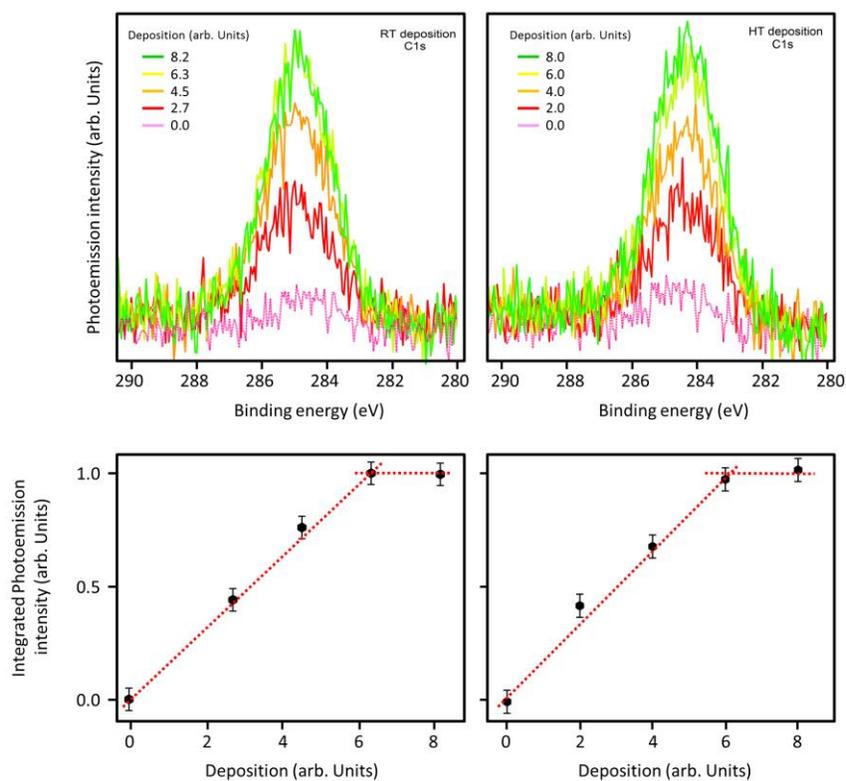

**Figure S1: Self-limited growth of supramolecular and polymeric phases.** (a,b) Evolution of the XPS spectra of the C1s core level as function of DBTP deposition onto a substrate kept at room temperature (a) and high temperature (b). (c,d) Evolution of the integrated signal as function of deposition, respectively associated to (a,b). The deposition is given by the evaporation time normalized with the molecular rate read on a quartz micro-balance.

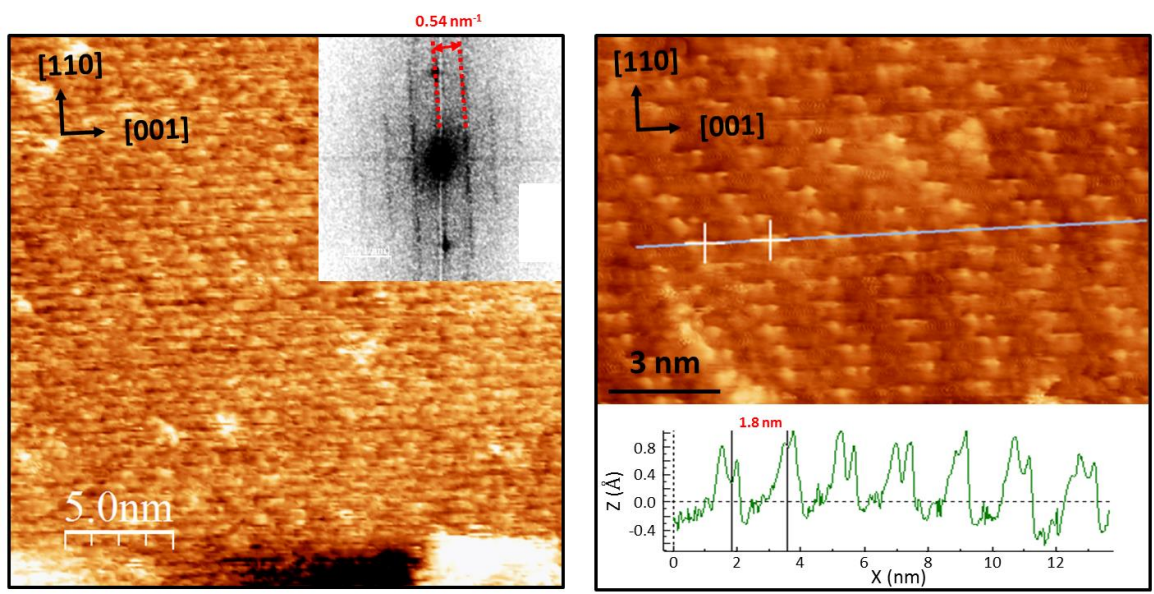

**Figure S2: Additional STM images of the supramolecular phase.** The inset in the left panel corresponds to the 2D Fourier transform of the image. The one in the right panel display a height profile corresponding to the grey line.

**Note 1: Geometry optimization of the DBTP molecule.**

The geometry optimization of the different molecular configurations was carried out using density functional theory (DFT), as implemented in the VASP code[1-3]. Inner electrons were described employing the projector augmented wave (PAW) method[4], and valence electrons were expanded in plane waves with an energy cutoff of 400 eV. The *optB88*-vdW functional[5], which accounts for non-local corrections, was adopted for the exchange and correlation potential. The Gamma point was selected for sampling the three-dimensional Brillouin zone. Isolated molecules as well as molecular dimers were fully relaxed, until residual forces were less than 0.02 eV/Å. According to this method, two energetically equivalent conformations of the free standing DBTP molecules were found, revealing a twisted conformation with C2h and D2 symmetry (Supplementary figure 2). In both cases, the twist angle (θ) is equal to 37.5° and the length (L) of the molecule, i.e. Br-Br distance, equal to 15.21 Å. The interaction between two molecules facing each other along the same axis was investigated for both conformations (Supplementary figure 3), revealing that supramolecular interactions stabilize the inter-molecular distance with a Br-Br distance of 3.8 Å.

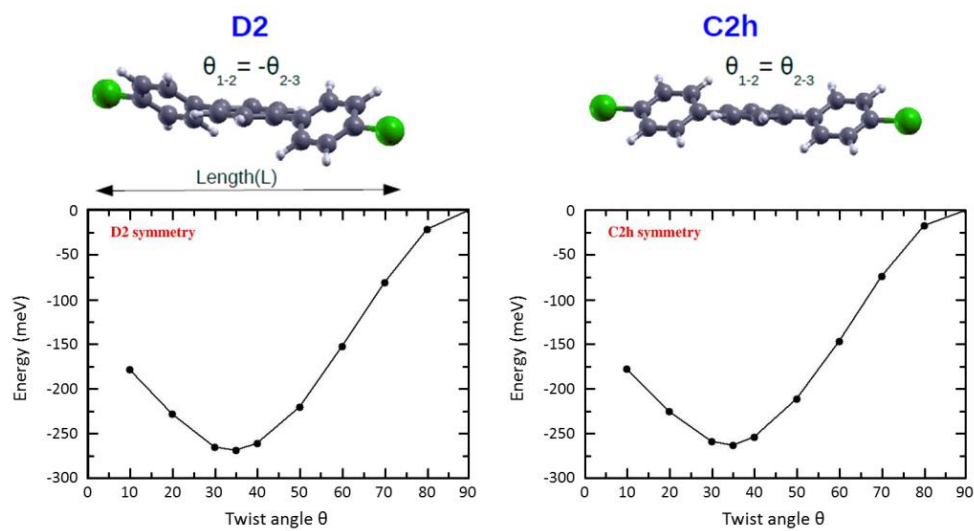

**Figure S3: Energy curves of the optimized molecule as function of the twist angle for the D2 and C2h conformations.** Left and right panels correspond respectively to the D2 and C2h symmetry cases.

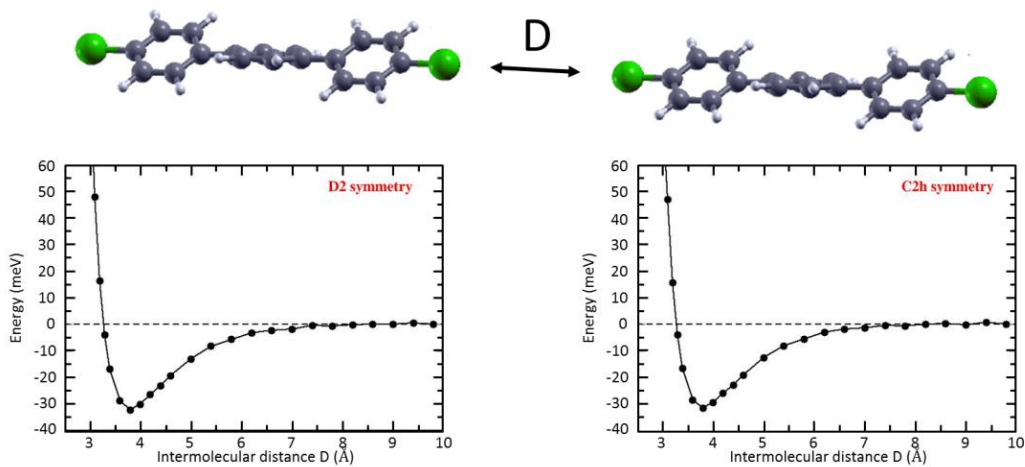

**Figure S4: Energy curves obtained for two facing molecules aligned along the same axis as function of the intermolecular Br-Br distance (D).** Left and right panels correspond respectively to the D2 and C2h symmetry cases.

**Note 2: Supramolecular and polymeric LEED patterns**

Supplementary figure 4 and 5 explain how the lack of phase relation between adjacent supramolecular and polymeric chains lead to the apparition of stripes running along the [-1,1,0] direction in LEED. In both cases, the reciprocal lattices corresponding to the different possible phases between neighboring chains are calculated. The sum of these different lattices give rise to a striped pattern, as the ones observed experimentally.

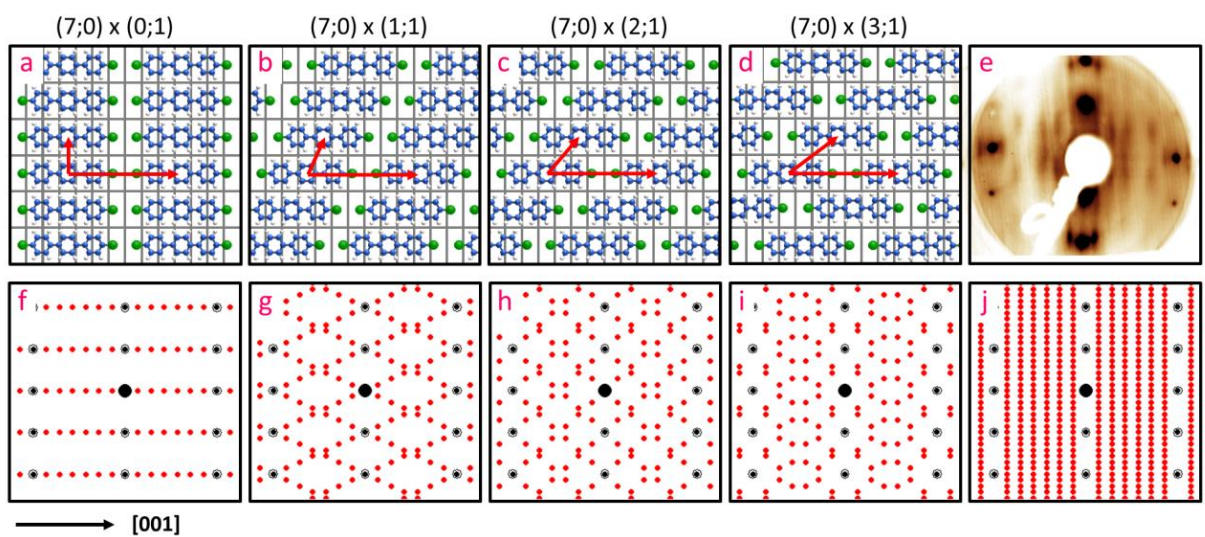

**Figure S5: Supramolecular LEED pattern.** (a-d) Schematic representations of four unit cells arising from different spatial shifts between neighbouring supramolecular DBTP chains. (f-i) Reciprocal lattices respectively corresponding to (a-d). The back spots correspond to the signature of the TiO$_2$ surface. (e) LEED pattern (45eV) measured for the supramolecular phase; (j) Superposition of the other four reciprocal lattices (f-i).

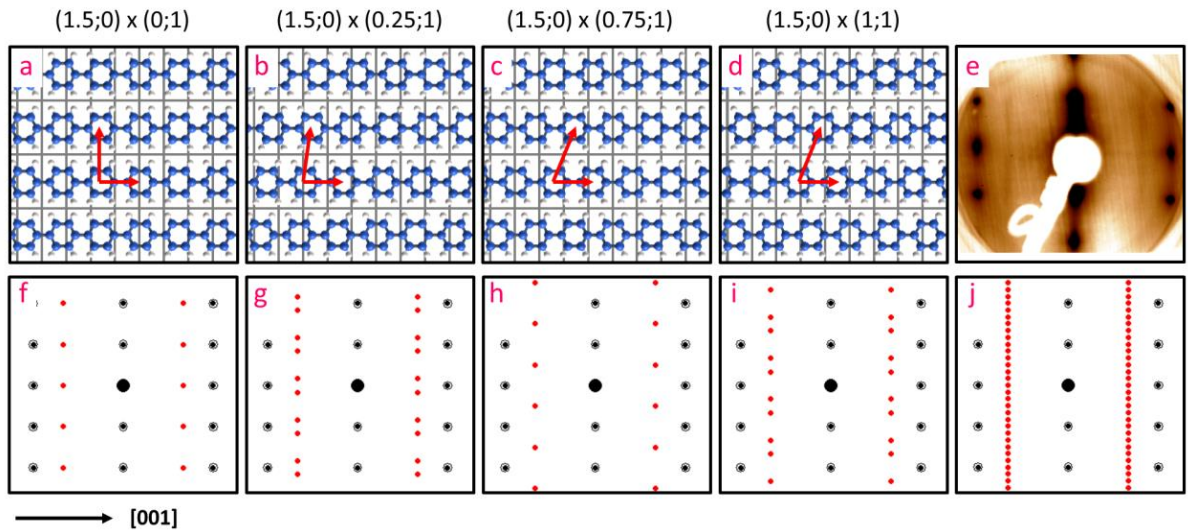

**Figure S6: Polymeric LEED pattern.** (a-d) Schematic representations of four unit cells arising from different spatial shifts between neighbouring PPP chains. (f-i) Reciprocal lattices respectively corresponding to (a-d). The black spots correspond to the signature of the TiO$_2$ surface. (e) LEED pattern (44eV) measured for the polymeric phase; (j) Superposition of the other four reciprocal lattices (f-i).

## Note 3: Phase transition in LEED

Supplementary Figure 6 presents the LEED patterns measured on a sample saturated with DBTP at room temperature and post-annealed to different temperatures. The signature of the supramolecular phase remains unchanged until 450K. At 490K, the supramolecular features vanish and only the spots of the clean surface are visible.

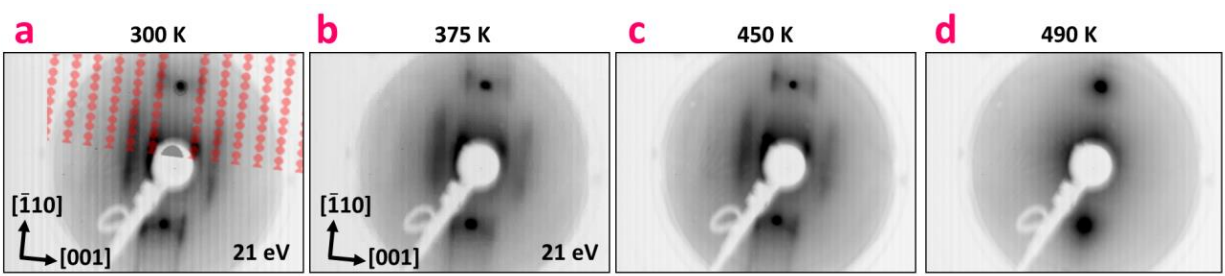

**Figure S7: Temperature phase transition of DBTP followed by LEED.** LEED patterns (21 eV) acquired for 1ML of DBTP evaporated on a sample held at RT (a) and post-annealed at 375 K (b), 450 K (c) and 490 K (d). The supramolecular reciprocal network deduced from figure S5 is overprinted on the left panel.

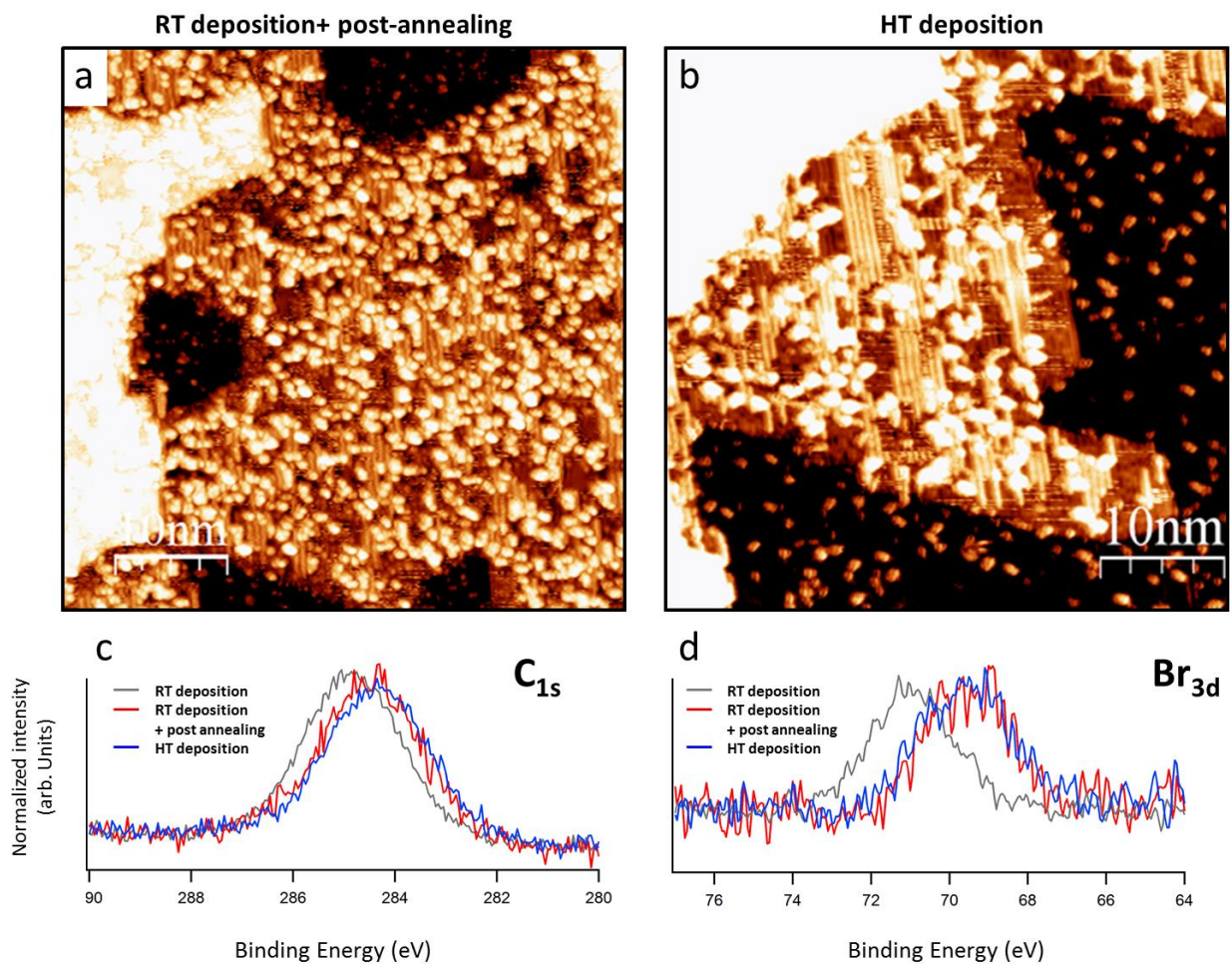

**Figure S8: Comparison between post-annealed room temperature preparation and high temperature deposition.** (a,b) STM images of the surface after (a) deposition of molecules at room temperature followed by post annealing over 520K, and (b) after deposition of molecules with the substrate kept at high temperature; (c,d) XPS measurements of C1s and Br3d core levels as function of preparation.

**Supplementary references.**